\documentclass[journal]{IEEEtran}
\IEEEoverridecommandlockouts

\usepackage[ruled,linesnumbered]{algorithm2e}
\usepackage{multirow}
\usepackage{makecell}
\usepackage{caption}
\usepackage{graphicx}  
\usepackage{float} 
\usepackage{subfigure}  
\usepackage{pdfpages}
\usepackage{tabularx}
\usepackage{amsmath}    
\usepackage{amsfonts,amssymb}
\usepackage{stmaryrd}
\usepackage{diagbox}
\usepackage[ruled]{algorithm2e}
\newtheorem{problem}{Problem}
\newtheorem{definition}{Definition}
\newtheorem{example}{Example}
\newtheorem{proposition}{Proposition}
\newtheorem{lemma}{Lemma}

\newtheorem{theorem}{Theorem}
\usepackage{cite}
\usepackage{bm}
\usepackage{xcolor}

\usepackage{CJK}
\usepackage{indentfirst}
\usepackage{amsmath}
\usepackage{cases}
\usepackage{subeqnarray}
\usepackage{booktabs}
\usepackage{bigstrut}
\usepackage{fancyhdr} 
\makeatletter

\newcommand{\Rmnum}[1]{\expandafter\@slowromancap\romannumeral #1@}
\makeatother
\title{\LARGE \bf Optimal Task and Motion Planning for Autonomous Systems  Using Petri Nets}

\author{Zhou He$^{1}$,~\IEEEmembership{Member,~IEEE}, Shilong Yuan$^{1}$,  
Ning  Ran$^{2}$, ~\IEEEmembership{Member,~IEEE}, 
and Dimitri Lefebvre$^{3}$, ~\IEEEmembership{Senior Member,~IEEE}
\thanks{$^{1}$Zhou He and Shilong Yuan are with the School of Electrical and Control Engineering, Shaanxi University of Science and Technology, Xi'an 710021, China (email: hezhou@sust.edu.cn; 220611020@sust.edu.cn)}%
\thanks{$^{2}$Ning Ran is with  Laboratory of Energy-Saving Technology,
College of Electronic Information Engineering, Hebei University, Baoding 071002, China (email: ranning87@hotmail.com)}%
\thanks{$^{3}$Dimitri Lefebvre is with the Universit\'e Le Havre Normandie, GREAH,  Le Havre 76600, France. (e-mail: dimitri.lefebvre@univ-lehavre.fr)}
}

\begin{document}
\maketitle
\thispagestyle{fancy} 
      \rhead{\thepage}
      \renewcommand{\headrulewidth}{0pt}
      \renewcommand{\footrulewidth}{0pt}
      \fancyfoot{}
\pagestyle{fancy}
\rhead{\thepage} 

\begin{abstract}
    This study deals with the problem of task and motion planning of {autonomous systems} within the context of high-level  tasks. Specifically, a task comprises logical requirements (conjunctions, disjunctions, and negations) on the trajectories and final states of agents in certain regions of interest.
   We propose an optimal planning approach that combines offline computation and online planning. First, a simplified Petri net system is proposed to model the autonomous system. Then, indicating places are designed to implement the logical requirements of the  specifications. Building upon this, a compact representation of the state space called extended basis reachability graph is constructed and an efficient online planning algorithm is developed to obtain the optimal plan. It is shown that the most burdensome part of the planning procedure may be removed offline, thanks to the construction of the extended basis reachability graph. 
   Finally, series of simulations are conducted to demonstrate the computational efficiency and scalability of our developed method.\\

\end{abstract}


\IEEEpeerreviewmaketitle

\section{Introduction}

\IEEEPARstart{T}{he}  use of autonomous systems, including unmanned ground vehicles (UGVs) and unmanned aerial vehicles (UAVs), has expanded to encompass various applications, such as  search and rescue operations, inspection tasks, and logistics and transportation. In certain applications of autonomous  systems, such as smart farms, smart factories, and other scenarios where environmental information is known in advance, { the main challenge is to rapidly identify an optimal solution to the task and motion planning (TAMP) problem.}

  {A growing number of researchers have directed their attention towards the problem of TAMP  for autonomous systems in recent years {\cite{TAMPRAL}}. The problem combines high-level task planning and low-level motion routing to produce workable solutions for finishing long-horizon activities, while the total cost of the system is optimized, e.g., total energy consumption, total task completion time, and so on \cite{survey2}.} 
A superior methodology for addressing TAMP in { autonomous systems} is to integrate the task planning with the motion planning process {\cite{TAMPRAL3,TAMPRAL2}}. This approach can yield more efficient and higher-quality solutions than decoupled methods. 
{Petri nets (PNs) are a useful tool for modeling and evaluating { autonomous systems} without enumerating the complete state space because of their compact representation of state space \cite{MRSs2}}.  



{
Recently, high-level tasks for { autonomous systems} have received much attention, where Boolean specifications \cite{Mahulea2017} and linear temporal logic (LTL)
{\cite{LTLRAL}} are commonly used as formal languages. Particularly, Boolean specifications are an effective ways of accurately expressing complex logical pertaining to safety, choices, endpoint requirements, and so on.   The TAMP for { autonomous systems} with Boolean specifications is investigated by several researchers. In \cite{Security}, { task planning} with Boolean specification and security requirements {for UGVs} is studied. In \cite{Probabilistic},  PNs are used to study the TAMP with Boolean specifications in unknown environments.
}



 This paper deals with the task and motion planning for TAMP problem with Boolean specifications using Petri nets.  
{The main contributions of this work are summarized as follows: (i) a state space reduction method for autonomous systems with task Boolean specifications is developed. By modeling autonomous systems with simplified PN systems and adding indicating places to implement the Boolean specification, a compact representation of the state space called \textit{extended basis reachability graph} (EBRG) is constructed offline}, which reduces the state space of the system significantly; (ii) an efficient searching algorithm for computing the optimal  task and motion planning of each agent is proposed based on the EBRG. Moreover, we prove that this plan preserves the optimality of the TAMP for autonomous systems with Boolean specifications; (iii) numerical experiments are investigated to illustrate the efficiency and scalability of the proposed approach when the number of agents and the size of the environment increase. Particularly, the results show that the developed approach is more scalable than the existing approaches.

In the rest of the paper,  section \ref{art} discusses the state of the art. 
The essential background is presented in Section \ref{Preliminary}, which includes the presentation of some basic concepts and problem formulation. Furthermore, the simplification of the model and the construction of the EBRG are described in Section \ref{3}. The methodology employed for the online search is elucidated in Section \ref{4}. The comparative results and simulation experiments are discussed in Section \ref{5}. Finally, the conclusion and future works are presented in Section \ref{conclusion}.

 \section{State of the art}\label{art}



{During the past decades, much attention has been paid to TAMP for autonomous systems with Boolean specifications. The goal of the problem is to plan paths for the whole team to fulfill the Boolean specifications while minimizing the total cost. To avoid the state explosion problem, the authors of \cite{Mahulea2017} model {autonomous systems} using PN systems. Some linearization methods are designed to represent the Boolean specification and trajectories of agents under the condition that the maximum number of movements is a predefined parameter.  This serves as the foundation for the presentation of an integer linear programming (ILP) approach that offers the best solution for TAMP problem with Boolean specifications. However, ILP is known to be an NP-hard problem, whose computational complexity increases significantly when the number of movements increases. 
The path planning problem for  autonomous systems with an unknown static environment is examined in \cite{Probabilistic}.
A heuristic method is developed that each agent solves an optimization problem and moves iteratively. In \cite{Zhe2024}, the problem of TAMP for { autonomous systems} 
with Boolean specification under motion and environment uncertainties is discussed. A Markov decision process model is constructed and an iterative reinforcement learning strategy 
is proposed. A meta-heuristic approach based on a simulated annealing algorithm (SAA) is suggested to find near-optimal  solutions for autonomous systems in order to minimize computational complexity \cite{simulated-annealing}. It should be highlighted, therefore, that the method proposed in \cite{simulated-annealing} does not ensure that the resultant solution is optimal.
}

 {It should be noted that the TAMP for autonomous systems with high-level tasks and security requirements has also received attention in recent few years \cite{Yin2023,YinCEP2022}, where the agents should not only complete complex tasks in accordance with Boolean specifications but also navigate paths that satisfy certain security requirements. The work of  \cite{Yuanjiaxing} defines a set of secret states for the system and requires that an outside intruder should never infer the secretes from finite observation of the system’s behaviors. An optimal approach based on PNs and ILP is developed to obtain a path for each agent that fulfills both the Boolean specifications and security constraints.}

 Our approach is different from the existing approaches, which
requires to solve an ILP problem \cite{Mahulea2017}. In\cite{Mahulea2017}, the authors assume that the maximal number of movements of the agents is a predefined parameter. Then, the motion strategy of each agent and the task specification are transformed into a series of linear constraints and an ILP problem is proposed to obtain an optimal or suboptimal solution. However, it is shown that the method cannot provide an optimal solution when the predefined parameter is set too small. On the contrary,  when this parameter is set too large, redundant constraints are introduced which result in high  computational complexity and low real-time performance. In comparison, our method provides an optimal solution since it builds the compact representation of the state space. Furthermore, once the offline construction of the basis reachability graph is done, the online planning only requires exploring a few nodes of the computed graph for a given task specification,  which makes our approach more efficient, in particular, for real time applications.

\section{Preliminary}\label{Preliminary}

\subsection{Petri net model for TAMP}\label{Petrimodel}	
		
A  Petri net model for TAMP (TAMP-PN) is an eight-tuple $Q=(P,T,$$Pre,Post,M_0, \bm{\Pi},h,w)$ \cite{Mahulea2017,LTL-BRG}, where $P$ is a finite set of $m$ \textit{places} denoted by circles; $T$ is a finite set of $n$ \emph{transitions} denoted by bars; {$Pre:P\times T\rightarrow \{0,1\}$ and $Post:P\times T\rightarrow \{0,1\}$} respectively are \emph{pre-} and \emph{post-incidence function} that connect places and transitions; $\bm{\Pi}$ is a set of atomic propositions; $h:P\rightarrow2^{\bm{\Pi}}$ is a labeling function that designates a set of atomic propositions to each place, where the set of atomic propositions that hold at place $p_i$ is represented by $h(p_i)$; 
$w:T\rightarrow\mathbb{R}^+$ is a cost function that assigns each transition $t\in T$ an integer to represent the cost spent to fire $t$; {$M:P\rightarrow\mathbb{N}$} is called a marking that is represented by an \emph{m-}component vector {$M \in \mathbb{N}^{m}$},  where $\mathbb{N}$ denotes  the set of non-negative integers. The count of tokens in place $p$ within marking {$M$} is represented by $M(p)$.  We use $M_0$ to denote the initial  distribution of agents
in the workspace.  We denote by $w=[w(t_1),w(t_2),...,w(t_n)]^T$ the cost vector.

Generally, each place denotes a region in the workspace, each transition denotes the movement from one region to another adjacent region, and each token denotes an agent. The \emph{incidence matrix} is defined by $C = Post - Pre \in \mathbb{Z}^{m\times n}$, where $\mathbb{Z}=\{0,\pm1,\pm2,...\}$. A transition $t$ is said to be \emph{enabled} at a marking {$M$ if $M \geq Pre(\cdot,t)$}, indicated as $M[t\rangle$. If $t$ is enabled at $M$, then it may fire and reach a new marking $M'=M+C (\cdot,t)$, which is denoted by $M[t\rangle M'$. The set of all finite sequences of transitions over $T$ is denoted by $T^*$. Then  $M[\sigma\rangle M'$ denotes that a transition sequence $\sigma=t_1t_2...\in T^*$ is enabled at $M$ and its occurrence yields $M'$. The vector $\mathbf{y}_\sigma$ is the \emph{firing vector} of $\sigma\in T^*$, i.e., $\mathbf{y}_\sigma(t)=r$ if transition $t$ occurs $r$ times in $\sigma$. For a transition $t\in T$, its \emph{input places} and \emph{output places} are indicated by $^{\bullet}t=\{p\in P| Pre(p,t)=1 \}$ and $t^{\bullet}=\{p\in P | Post(p,t)=1\}$, respectively.
An analogous definition is given to the input transitions $^{\bullet}p$ and the output transitions $p^{\bullet}$ of place $p\in P$.
We denote by $P_0=\{p|M_0(p)>0\}$ the set of non-empty places at initial marking $M_0$ and  by $\mathcal{P}_f=\{\pi | h(p)=\pi\in \bm{\Pi},~\forall p \in P\}$ the set of propositions corresponding to possible final tasks. 
 We denote by $R(Q)$ the set of  all \textit{reachable markings} from  $M_0$ in $Q$.  

\subsection{Task specification}\label{Boolean}

 We denote by $\bm{\Lambda}=\{\Lambda_1,\Lambda_2,...\}$ the set of regions of interest; $\bm{\Pi}=\{\Pi_1,\Pi_2,...,\pi_1,\pi_2,...,\}$ the set of atomic propositions; and   $\delta:\Lambda\rightarrow 2^{\bm\Pi}$ a labeling function that assigns a set of atomic propositions to each region of interest. In particular, each proposition $\Pi_i$ (resp., $\pi_i$) indicates that  the region corresponding to $\Pi_i$ (resp., $\pi_i$) should be visited along the trajectory by an agent (resp., should be visited  at the final state by an agent. 

Syntactically, the specification of the { autonomous systems} is given as a Boolean formula:
\begin{equation}
\varphi=\varphi_I\wedge\varphi_D\wedge\varphi_F.  
\end{equation}
Particularly, subspecification $\varphi_I$=$\varphi_{I,1}\wedge\cdots\wedge\varphi_{I,a}$ indicates the requirements for the intermediate tasks that need to be completed by agents as they follow their trajectories, where each term
$\varphi_{I,j}=\Pi_{I,j1}\vee\Pi_{I,j2}\vee\cdots$ define the possible atomic propositions for term $\varphi_{I,j}$. Subspecification $\varphi_D$=$\varphi_{D,1}\wedge\cdots\wedge\varphi_{D,b}$ indicates the requirements for the last tasks that at least one agent must perform, where each term
$\varphi_{D,j}=\pi_{D,j1}\vee\pi_{D,j2}\vee\cdots$ define the possible atomic propositions for term $\varphi_{D,j}$.
Subspecification {$\varphi_F=[\neg\Pi_{F,1}|\neg\pi_{F,1}]\wedge[\neg\Pi_{F,2}|\neg\pi_{F,2}]\wedge\cdots$ denotes the requirements on regions that should be avoided, where each term $\neg\Pi_{F,i}$ (resp., $\neg\pi_{F,i}$) indicates that $\Lambda_j$ should be avoided along all trajectories, where $\delta(\Lambda_j)=\Pi_{F,i}$ (resp., all agents should not stay in region $\Lambda_j$ at the final state, where $\delta(\Lambda_j)=\pi_{F,i}$)}, square brackets  ``$[...]$'' indicate optional propositions, while the choice between two propositions is represented by ``$|$''. 
  


\begin{figure}[htbp]
\centering
\subfigure[An autonomous system]{\includegraphics[scale=0.35]{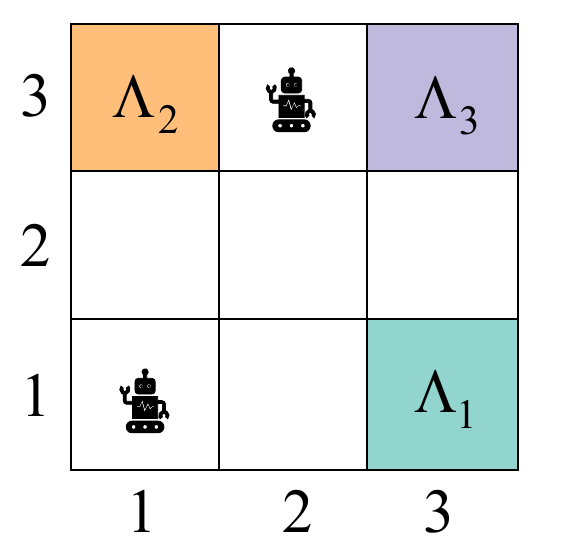}}\hspace{1em}
\subfigure[The TAMP-PN model $Q$]{\includegraphics[scale=0.25]{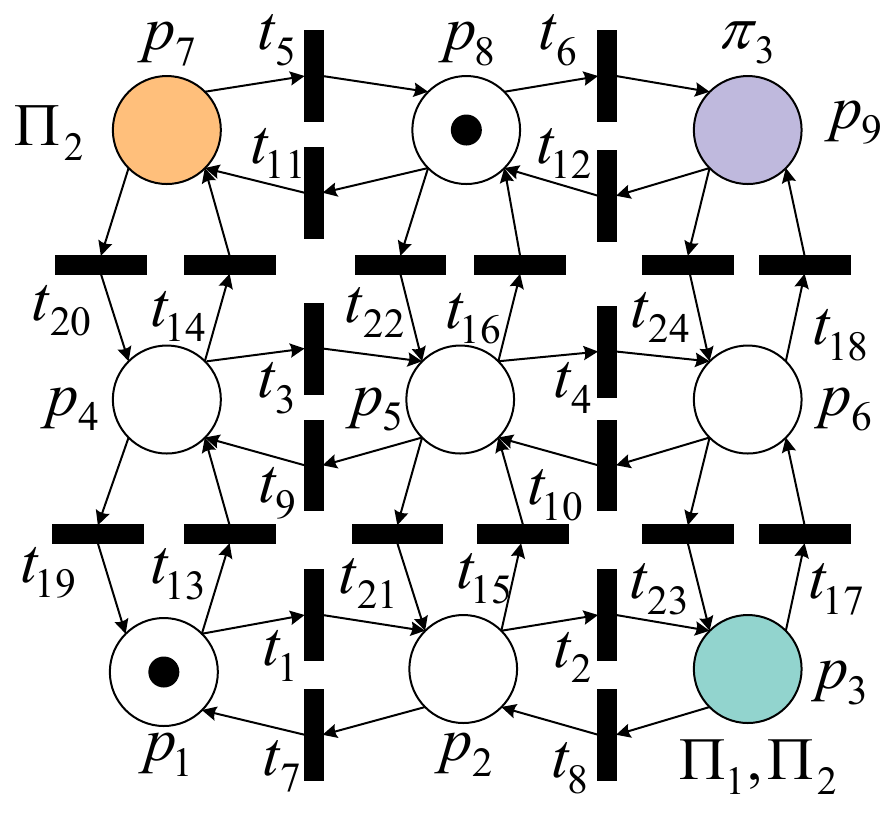}}
\caption{PN model for TAMP.}
\label{fig}
\end{figure}

\begin{example}\label{example1}
    Let us consider an autonomous system in Fig. 1(a) that consists of two agents working in a known environment containing nine regions.  The set of regions of interest is $\bm{\Lambda}=\{\Lambda_1,\Lambda_2,\Lambda_3\}$,  the set of atomic proposition is $\bm{\Pi}=\{\Pi_1, \Pi_2, \Pi_3, \pi_1, \pi_2, \pi_3\}$, and the labels are $\delta(\Lambda_1)=\{\Pi_1,\Pi_2\}$, $\delta(\Lambda_2)=\{\Pi_2\}$, and $\delta(\Lambda_3)=\{\pi_3\}$. The task specification $\varphi=\Pi_2\wedge\pi_3\wedge\neg\Pi_1$ which requires that either region $\Lambda_1$ or region $\Lambda_2$ should be visited along the trajectory, region $\Lambda_3$ must accessed by one agent in the final state, and region $\Lambda_1$ should be avoided along all trajectories. We assume that the agents can only move up/down/left/right to adjacent regions with one unit cost (i.e., $w = \vec{1}$). The TAMP-PN model is shown in Fig. 1(b), where $P=\{p_1,\ldots,p_9\}$, $T=\{t_1,\ldots,t_{24}\}$, $M_0=[1,0,0,0,0,0,0,1,0]$,  $h(p_3)=\{\Pi_1,\Pi_2\}$, $h(p_7)=\{\Pi_2\}$, $h(p_9)=\{\pi_3\}$, and $h(p)=\varepsilon$, $\forall p \in P \setminus \{p_3,p_7,p_9\}$. 
\end{example}

\subsection{Problem formulation}\label{Problem}


{ As stated in subsection \ref{Petrimodel}}, a TAMP problem can be described by a TAMP-PN model $Q$ and its  solution can be represented by a transition sequence  $\sigma=t_0t_1\cdots t_n$. { Therefore the total cost of all agent movements can be expressed as the cost corresponding to the transition sequence.} We call the transition sequence  $\sigma$ a \textit{run} of $Q$. 
{ We say  that  $\sigma$ is a feasible run (i.e., solution)  satisfying $\varphi$ if $h(t_0^{\bullet})h(t_1^{\bullet})\cdots h(t_n^{\bullet})$ satisfies $\varphi$,  denoted by $\sigma\vDash\varphi$.}  
~Let $\Sigma(Q)$ be the set of all feasible transition sequences and $\Sigma_\varphi(Q)$ be the set of all feasible transition sequences satisfying $\varphi$. The problem of task and motion planning for autonomous systems can be formulated as follows.   

\begin{problem}(TAMP)\label{problem1}
Given a Boolean specification $\varphi$ and a TAMP-PN $Q$ representing a team of homogeneous agents working in a static and known environment, determine an optimal sequence $\sigma^\star$ for the team such that
	\begin{itemize}
		\item $\sigma^\star\in\Sigma_\varphi(Q)$,
		\item $\forall\sigma\in\Sigma_\varphi(Q):w^T\cdot \mathbf{y}_{\sigma^\star}\leq w^T\cdot \mathbf{y}_{\sigma}$ , where $w^T\cdot \mathbf{y}_{\sigma^\star}$ is the cost of ${\sigma^\star}$.
        \end{itemize}
\end{problem}

\section{Model simplification}\label{3}


The workspace of a TAMP problem can be modeled as a TAMP-PN after discretization. However, the state space of the system increases significantly with respect to the size of the environment (i.e., the number of regions), which poses serious challenges for the search of optimal planning. In the following, we propose a method to simplify the TAMP-PN  to reduce the state space of the system.





The satisfaction of the Boolean specification is contingent upon the  visit of the propositional regions (i.e., places with non-empty propositions). Considering that the places associated with the agent's initial positions (i.e., non-empty places at initial marking) should also be kept, we preserve and abstract the minimal cost transition sequence between each place into a unique transition based on an ILP. Therefore, The simplified PN model resulting from this approach contains fewer places and transitions compared to the initial model $Q$, and does not lose any information regarding atomic propositions.


Given any $p\in P_0 \cup P_\Pi$ and $p'\in P_\Pi$, where $P_0$ is the set of non-empty places at
initial marking and $P_\Pi$ is the set of places  with  non-empty propositions, the set of all sequences that starting from $p$ to $p'$ without passing through any other place $p\in P_\Pi$
is denoted by $\sigma_{pp'}=\{\sigma \in T^*|^{\bullet}\sigma_0=p\wedge\sigma_{|\sigma|}^{\bullet}=p'\wedge(\forall 0\leq i\textless |\sigma|)[{\sigma}^{\bullet}_i \notin P_{\Pi}\wedge{\sigma}^{\bullet}_i={^{\bullet}\sigma_{i+1}}]\}$, and the minimal cost of sequences in $\sigma_{pp'}$ is denoted by $\hat{w}_{pp'}$, i.e.,
$\hat{w}_{pp'}=\mathop{min}\limits_{\sigma\in\sigma_{pp'}} w^{T}\cdot\mathbf{y}_\sigma$. 
{If $p' \in P_\Pi$ is not reachable from $p$, then  $\hat{w}_{pp'}=\infty$.   We suppose that there exists a single sequence $\hat{\sigma}_{pp'}$, called \textit{minimal sequence}, with the lowest cost $\hat{w}_{pp'}$ when $\hat{w}_{pp'}\neq\infty$. 
 {We denote by  $\hat{\Sigma}=\{\hat{\sigma}_{pp'}\in T^*|\exists p \in P_0\cup P_\Pi, p'\in P_\Pi~s.t. ~\hat{w}_{pp'}\neq\infty\}$ the set of all minimal sequences in $Q$.}} 
	
	The minimal sequence $\hat{\sigma}_{pp'}$ from $p\in P_0 \cup P_\Pi$  to $p'\in P_\Pi$ can be equivalently formulated as {the minimal cost accessibility problem of a token moving from $p$ to $p'$. Let $M_p$ (resp., $M_{p'}$) be a marking that only contains 1 token in place $p$ (resp., $p'$) and all other places are empty, 
the minimal sequence ${\hat{\sigma}_{pp'}}$ can be computed  by solving following ILP:
	\begin{equation}
		\begin{aligned}
			\text{min} \quad & w^{T} \cdot \mathbf{y}_{\hat{\sigma}_{pp'}} \\
			\text{s.t.} \quad & M_{p'} = M_{p} + C \cdot \mathbf{y}_{\hat{\sigma}_{pp'}} \\
			& Post(p'',\cdot) \cdot \mathbf{y}_{\hat{\sigma}_{pp'}} \leq 0, \forall p'' \in P_\Pi\setminus{\{p,p'\}} \\
			& \mathbf{y}_{\hat{\sigma}_{pp'}}\in \mathbb{N}^{n}
		\end{aligned}
	\end{equation}
where 	$Post(p'',\cdot)$ represents the  row of the
matrix $Post$ associated with place $p''$.} 
Note that ILP (2) has $(m+|P_{\Pi}|-2)$ constraints and $n$ variables. To simplify the model $Q$, it is required to determine the minimal sequence between every pair of the two places in $P_\Pi$, as well as from every place in $P_0$ to every place in $P_\Pi$. 

	
We create a new transition by abstracting each sequence $\hat{\sigma}_{pp'}\in\Sigma$.  The set of all abstracted transitions is represented by $T^s$, while the set of all abstracted locations is represented by $P^s= P_0\cup P_\Pi$. The bijection mapping from $T^s$ to $\hat{\Sigma}$ is denoted by $f: T^s\rightarrow \hat{\Sigma}$. Given a Petri net model $Q$, based on the set $T^s$ and $P^s$, the  \emph {simplified Petri net model} $Q^s$ is defined as follows.  
	
	
\begin{definition}	\label{Qs}
{(\textit{Simplified TAMP-PN}) Given a TAMP-PN $Q=(P,T,Pre,Post,M_0,\bm{\Pi},h,w)$, the abstracted transition set $T^s$, the projection function $f$, the simplified TAMP-PN is defined as an eight-tuple $Q^s=(P^s, T^s, Pre^s, Post^s$, $M^s_0,\bm{\Pi},h^s,w^s)$, where
		\begin{itemize}
                \item $P^s=P_0\cup P_\Pi$ is the set of places;
			\item $T^s$ is the set of transitions;
			\item $Pre^s$ (resp., $Post^s$): $P^s\times T^s \rightarrow \{0,1\} $ is the pre-(resp., post)-incidence function, which is defined as follows by $t^\bullet$ and $^\bullet t$:\\ 
                -~$|^{\bullet}t|=|t^\bullet|=1$, $\forall t \in T^s$;\\
			-~$[(p_i=^\bullet t)\wedge (p_j=t^\bullet)]\Rightarrow [f(t)=\hat{\sigma}_{p_ip_j}]$, $\forall t \in T^s$.
			\item $M^s_0$ is the initial marking such that:\\
			-~$M^s_0(p)=M_0(p) $, $\forall p \in P_0$;\\
			-~$M^s_0(p)=0$, $\forall p \in P_\Pi$;
			\item $\bm{\Pi}$ is the set of all atomic propositions;
            \item $h^s: P^s\rightarrow 2^{\bm{\Pi}}$ is the labeling function; 
			\item $w^s: T^s\rightarrow \mathbb{R}^+ $ is the cost function that gives each transition $t\in T^s$ a cost such that: $\forall t \in T^s$, $[(w^s(t)=\hat{w}_{p_ip_j})\wedge (p_i={^\bullet t})\wedge (p_j=t^\bullet)]$.
		\end{itemize}
	}
	\end{definition}

{Let $\mathcal{F}:\Sigma(Q^s)\rightarrow\Sigma(Q)$ be a function  such that: $\forall\sigma=t_0t_1t_2\cdots$, $\sigma\in\Sigma(Q^s)$, $\mathcal{F}(\sigma)=f(t_0)f(t_1)f(t_2)\cdots$, $\mathcal{F}(\sigma)\in\Sigma(Q)$.} Lemma 1 demonstrates how the projection of the optimal sequence in $Q^s$ could provide the optimal sequence for Problem 1.

{\begin{lemma}\label{lemma1}
     There exists a solution $\sigma^\star$ to Problem \ref{problem1} iff there exists a sequence $\sigma_s^\star$ in $Q^s$ such that
     \begin{itemize}
         \item ${\sigma}_s^{\star}\in\Sigma_\varphi(Q^s);$
         \item $\forall\sigma_s\in\Sigma_\varphi(Q^s):(w^s)^T\cdot\mathbf{y}_{\sigma_s^\star}\leq (w^s)^T\cdot \mathbf{y}_{\sigma_s}$;
          \item
$(w^s)^T\cdot\mathbf{y}_{\sigma_s^\star}=w^T\cdot\mathbf{y}_{\mathcal{F}({\sigma}_s^\star)}=w^T\cdot\mathbf{y}_{\sigma^\star}$.       
     \end{itemize} 
 \end{lemma}}

\textit{Proof: (If)}  According to the function $\mathcal{F}:\Sigma(Q^s)\rightarrow\Sigma(Q)$ and the definition of $w^s$, if there exists a sequence $\sigma^\star_s$ with the minimal cost that satisfies $\varphi$, there exists an equivalent sequence $\sigma^\star$ in $Q$ such that $\sigma^\star\vDash\varphi$ has the minimal cost. Therefore, $\sigma^\star$ is a solution to Problem \ref{problem1}. 

\textit{(Only if)} According to Problem \ref{problem1}, if there exists a solution $\sigma^\star$, then $\sigma^\star$ satisfies the specification $\varphi$ and has the minimal cost. Based on the definition of transition set $T^s$ and $w^s$, there exists an equivalent sequence $\sigma^\star_s$ in $Q^s$ such that $\sigma^\star_s\vDash\varphi$ has the minimal cost.

\section{Planning procedure} \label{4}
 \subsection {Monitored PN model}\label{controller}
 Inspired by the works in \cite{Controlplace1}, we introduce indicating places into the simplified PN model $Q^s$ to further implement the Boolean specifications. 
Such places capture the information on whether the atomic propositions have been satisfied or not.   
For each proposition $\Pi_i\in \mathcal{P}_t$, we design an indicating places $p_i^{obs}$ that connects all the output transitions of places $p$ which are labelled with  $\Pi_i$, i.e.,  $h^s(p)=\Pi_i$. Specifically, for each transition $t\in {p}^\bullet$, where $h^s(p)=\Pi_i$, we have $Pre^m(p_i^{ind},t)=1$. 
Therefore, there exist  $r$ indicating places in total.  We denote by { $P^{ind}=\{p^{ind}_1,p^{ind}_2,...,p^{ind}_{r}\}$} the set of indicating places. For the remainder of the paper, we assume that an indicating place can have a maximum of one token.

 \begin{definition}\label{Qc}
 (\textit{Monitored TAMP-PN}). Given a simplified TAMP-PN $Q^s=(P^s,T^s,Pre^s,Post^s,M^s_0$, $\bm{\Pi},h^s,w^s)$, the monitored TAMP-PN is defined as an eight-tuple $Q^m=(P^m,T^m,Pre^m,Post^m, M^m_0,\bm{\Pi},h^m,w^m)$, where
		\begin{itemize}
			\item $P^m=P^s\cup P^{ind}$ is the set of places;
			\item $T^m=T^s$ is the set of transitions;
{			\item $Pre^m$ (resp., $Post^m$): $P^m\times T^m\rightarrow\{0,1\}$ is the pre (resp., post)-incidence function; 
                \item $M^{m}_0\in \mathbb{N}^{|P^s+r|}_{\geq 0}$ is the initial marking such that:\\
			-~$M^m_0(p)=M^s_0(p)$, $\forall p \in P_0$;\\
			-~$M^m_0(p)=0$, $\forall p \in P_\Pi\cup P^{ind}$;
                \item $h^m:P^m\rightarrow 2^{\bm{\Pi}}$ is the labeling function, where $h^m(p)=\varepsilon$, $\forall p \in P^{ind}$;
                \item $w^m=w^s$ is the cost function.} 
		\end{itemize}

  Compared with the TAMP-PN $Q$ that has $m$ places, the monitored TAMP-PN only requires $|P_0|+|P_{\Pi}|+r$ places to model the same autonomous system, which is usually much less than $m$. Moreover, the fulfillment of the specification $\varphi$ can be detected by the indicating places as follows:
\begin{itemize}
		\item	negation: $\varphi=\neg\Pi_i$ (resp., $\varphi=\neg\pi_i$) is equivalent to $M(p^{ind}_i)=0$ (resp.,  for each place $p$ with $h^m(p)=\Pi_i$, it holds $M(p)=0$); 
		\item   conjunction: $\varphi=\Pi_i\wedge\Pi_j$ (resp., $\varphi=\pi_i\wedge\pi_j$) is equivalent to $M(p^{ind}_i)=1$ and $M(p^{ind}_j)=1$ (resp., 
 for each place $p$ with $h^m(p)\in \{\Pi_i,\Pi_j\}$, it holds $M(p)\geq 1$);
		\item  disjunction: $\varphi=\Pi_i\vee\Pi_j$ (resp., $\varphi=\pi_i\vee\pi_j$) is equivalent to $M(p^{ind}_i)+M(p^{ind}_j)\geq 1$ (resp.,  there exists a place $p$ with $h^m(p)\in \{\Pi_i,\Pi_j\}$ such that $M(p)\ge1$).
	\end{itemize}

{ By Definition \ref{Qc}, since $T^m=T^s$, it is clear that there also exists the function $\mathcal{F}$ between $\Sigma(Q^m)$ and $\Sigma(Q)$ such that $\forall\sigma=t_0t_1t_2\cdots,\sigma\in\Sigma(Q^m)$, $\mathcal{F}(\sigma)=f(t_0)f(t_1)f(t_2)\cdots,\mathcal{F}(\sigma)\in\Sigma(Q)$.}
 In Lemma \ref{lemma2},  we show that the optimal sequence in $Q^s$ can be equivalently obtained by the projection of the optimal sequence in $Q^m$.
\end{definition}

\begin{lemma}\label{lemma2}
    There exists a solution $\sigma_s^\star$ in $Q^s$ iff there exists a sequence $\sigma_m^\star$ in $Q^m$ such that
    \begin{itemize}
    \item ${\sigma}_m^\star={\sigma}_s^\star;$
    \item $(w^m)^T\cdot\mathbf{y}_{{\sigma}_m^\star}=(w^s)^T\cdot\mathbf{y}_{{\sigma}_s^\star}.$
    \end{itemize} 
\end{lemma}

\textit{Proof: (If)} Since the transition set $T^m$ is equivalent to $T^s$ {(i.e., $T^s=T^m$)} and the added indicating places have no impact on the firing of transitions in $T^s$, if there exists a transition sequence $\sigma_m^\star$ in $Q^m$, we can find an equivalent transition sequence $\sigma_s^\star=\sigma_m^\star$ in $Q^s$. Moreover, since $w^m=w^s$, we have  $(w^m)^T\cdot\mathbf{y}_{{\sigma}_m^\star}=(w^s)^T\cdot\mathbf{y}_{{\sigma}_s^\star}$.

\textit{(Only if)} { Since $T^s$ is equivalent to $T^m$ and $w^s=w^m$, if there exists a transition sequence $\sigma_s^\star$ in $Q^s$, we can find an equivalent transition sequence $\sigma_m^\star=\sigma_s^\star$ in $Q^m$ such that  $(w^s)^T\cdot\mathbf{y}_{{\sigma}_s^\star}=(w^m)^T\cdot\mathbf{y}_{{\sigma}_m^\star}$.}
\hfill  $\blacksquare$



     It is obvious that if $\sigma_m^\star$ is the optimal sequence to Problem \ref{problem1} in $Q^m$, then $\mathcal{F'}(\sigma_m^\star)$ is an optimal solution to Problem \ref{problem1}.


\subsection{Off-line computation of extended basis reachability graph}\label{brg}

 A compact structure of state space of PN  called basis reachibility graph (BRG) is proposed in \cite{BRG3},
based on which the minimal sequence search
problem can be solved more efficiently.  An extended BRG (EBRG) in this paper is proposed to improve the trajectory search efficiency. The main difference  with BRG is that only the path with minimal cost between each state is preserved in EBRG, while all the feasible sequences are considered in the original BRG. Thus, the size of the EBRG in this paper is smaller than BRG \cite{BRG3}, which can further reduce the searching complexity. In the following, we recall some related concepts of  BRG. For more details, we refer  to \cite{BRG3}. 

\begin{definition}\label{definition3} 
{(\textit{Basis partition of transitions}) \cite{BRG3} Given a PN $N$,  a \emph{basis partition} of transition set $T$ is a pair $\varpi=(T_E,T_I)$ such that: (i) $T_I\subseteq T$, $T_E=T\setminus  T_I$; and (ii) the $T_I$-induced subnet is acyclic. The sets $T_E$ and $T_I$ are said to be the sets of explicit transitions and  implicit transitions, respectively.}
\end{definition}

\begin{definition} \label{definition4}
(\textit{Explanations}) \cite{BRG3} Given a PN $N$, a basis partition $\varpi=(T_E,T_I)$, a marking $M$, and a transition $t\in T_E$, we define:
	\begin{itemize}
		\item $\Sigma(M,t)=\{\sigma\in T^*_I|M[\sigma\rangle M', M'\geq Pre(\cdot,t)\}$ as the set of \emph{explanations} of transition $t$ at marking $M$;
            \item $\Sigma_{min}(M,t)=
            \{\sigma\in\Sigma(M,t)|\nexists\sigma'\in\Sigma(M,t), \mathbf{y}_{\sigma'}\lneq \mathbf{y}_{\sigma}\}$ as the  set of \textit{minimal explanations};
		\item $Y_{min}(M,t)=\{\mathbf{y}_\sigma\in \mathbb{N}^{n}|\sigma\in \Sigma_{min}(M,t)\}$ as the set of \emph{minimal explanation vectors}.
        \end{itemize}
\end{definition}

Note that $\Sigma(M,t)$ denotes  the set of  implicit transition sequences that should  fire at marking $M$ to enable an explicit transition $t$,  $\Sigma_{min}(M,t)$ denotes the set of sequences in $\Sigma(M,t)$ with minimal firing sequences, and $Y_{min}(M,t)$ denotes the set of all minimal elements of $Y(M,t)$.  

%

%
\begin{definition}\label{BRGB} (Basis reachability graph) \cite{BRG3}
    Given a PN $N=(P,T,Pre,Post)$ with an initial marking $M_0$ and a basis partition $\varpi=(T_E,T_I)$, its \emph{basis reachability graph} (BRG) is a four-tuple $B=\{\mathcal{M}_{basis},Tr,\Delta,M_0\}$, where
	\begin{itemize}
		\item $\mathcal{M}_{basis}$ is the set of basis markings; 
		\item $Tr$ is the set of pairs $(t,\mathbf{y})\in T_E\times \mathbb{N}^{n}$;
		\item $\Delta\subseteq\mathcal{M}_{basis}\times Tr\times\mathcal{M}_{basis}$ is a transition relation such that {$\Delta=\{(M,(t,\mathbf{y}),M')|\mathbf{y}\in Y_{min}(M,t),t\in T_E,M'=M+C\cdot\mathbf{y}+C(\cdot,t)\}$;}
		\item $M_0\in \mathcal{M}_{basis}$ is the initial marking.
	\end{itemize}
\end{definition}%

 Note that in order to find the minimal sequence for Problem 1, we propose an EBRG $\mathcal{B}=\{\mathcal{M}_{basis},Tr_q,\Delta_q,M_0\}$ in Algorithm~1, where the transition relation is $\Delta_q=\{(M,(t,\mathbf{y}),M'),q(M')\}$
and $q(M')\in\mathbb{R}^+$ denotes the accumulated cost from $M_0$ to $M'$ by firing the transition sequence $\sigma=\mathbf{y}t$. Moreover, we only preserve the path with the minimal cost $q$ between each basis marking pair, resulting in $Tr_q\subseteq Tr$. Consequently, the number of edges and vertices in the EBRG $\mathcal{B}$ is $|\mathcal{M}_{basis}|-1$ and $|\mathcal{M}_{basis}|$, respectively.

According to Definition \ref{BRGB}, an monitored TAMP-PN $Q^m$ can have multiple EBRG depending on different basis partitions $\varpi=(T_E,T_I)$. So far, there is no quantitative relationship between the size of the BRG and the choice of basis partition $\varpi$. In this paper, we use the partitioning method in \cite{BRG3} to obtain a good basis partition $\varpi=(T_E,T_I)$ to reduce the size of the EBRG as much as possible. Moreover, for the places that are labeled with propositions $\pi_i \in \mathcal{P}_f$, their input transitions are preserved in the set of explicit transitions $T_E$.
\begin{algorithm}[!htpb]\scriptsize
	\SetAlgoLined 
	\caption{Extended basis reachability graph}
	\KwIn{$Q^m=(P^m,T^m,Pre^m,Post^m,M^m_0,\bm{\Pi},h^m,w^m)$ }
	\KwOut{EBRG $\mathcal{B}=\{\mathcal{M}_{basis},Tr_q,\Delta_q,M_0\}$}
	Select a basic partition $\varpi=(T_E,T_I)$\\
	Let $\mathcal{M}_{basis}=\emptyset$,  $\mathcal{M}_{new}=\{M^m_0\}$\\ 
	\While {\textnormal{$\mathcal{M}_{new}\neq\emptyset$}}{
		\For{each $M\in\mathcal{M}_{new}$}
		{
			\For{each $t\in T_E$}
			{
				\For{each $\mathbf{y}\in Y_{min}(M,t)$}
				{
					$M'\leftarrow M+C\cdot \mathbf{y}+C(\cdot,t)$\\
					{ $q(M')\leftarrow q(M)+(w^m)^T\cdot \mathbf{y}+w^m(t)$\\}
					\If{$\nexists M'\in\mathcal{M}_{basis}\cup\mathcal{M}_{new}$ }
					{ 
						 $\mathcal{M}_{new}\leftarrow\mathcal{M}_{new}\cup\{M'\}$\\
						{ $\Delta_q\leftarrow\Delta_q\cup(M,(t,\mathbf{y}),M',q(M'))$\\}
						
					{
							\ElseIf {there exists { $(\cdot,(t'',\mathbf{y}''),M'',q(M''))\in\Delta_q$ with $M''=M'$ and $q(M'')>q(M')$}}
							{
								{ Updte $(\cdot,(t'',\mathbf{y}''),M'',q(M''))$ to $(M,(t,\mathbf{y}),M'',q(M'))$\\}
							}
						}
					}
				}
			}
		}
        $\mathcal{M}_{basis}\leftarrow\mathcal{M}_{basis}\cup\{M\}$\\
        $\mathcal{M}_{new}\leftarrow\mathcal{M}_{new}\setminus\{M\}$\\
	}
	\textbf{return} EBRG $\mathcal{B}$\\
\end{algorithm}	
 

 	 \begin{proposition}\label{basissequence}
	    	 \cite{BRG3} Given a PN system $Q$,  its EBRG $\mathcal{B}=(\mathcal{M}_{basis},Tr,\Delta_q,M_0)$ corresponding to $\varpi=(T_E,T_I)$, and a marking $M$, the following two statements are equivalent:
		\begin{itemize}
			\item There exists a  transition sequence $\sigma=\sigma_1t_{1}\cdots\sigma_nt_{n}\sigma_{n+1}$ ($\sigma_i\in T^*_I$ and $t_i\in T_E$) such that $M_0[\sigma\rangle M$;
			\item There exists a {path} in the EBRG $\mathcal{B}$ 
             \begin{small}
			\begin{equation}
{ M_0\xrightarrow{(t_{1},\mathbf{y}_1),q(M_1)} M_{1}\xrightarrow{(t_{2},\mathbf{y}_2),q(M_2)}\cdots\xrightarrow{(t_{n},\mathbf{y}_n),q(M_n)}M_{n}}
			\end{equation}            			
            \end{small}
such that $M\in \{M_{n+1}|M_{n}[\sigma'\rangle M_{n+1},\sigma'\in T^*_I\}$.            
	\end{itemize}
	\end{proposition}

{

 If there exists a path in the EBRG $\mathcal{B}$ that has a minimal cost and satisfies the given Boolean specification $\varphi$, then 
 there exists an optimal sequence $\sigma_m^\star$ in $Q^m$ to Problem 1.

 \subsection{On-line planning strategy}

{

Based on the EBRG $\mathcal{B}$, we aim to find a minimal path starting from the initial marking $M_0$ to a target marking $M_{tg}\in \mathcal{M}_{basis}$ such that the Boolean specification $\varphi$ is fulfilled at marking $M_{tg}$ (i.e., the path satisfies $\varphi$). In the following, we provide an  ILP to find the target marking $M_{tg}$ in the EBRG $\mathcal{B}$.
}

For each proposition $\varphi_{I,i}$ of  $\varphi_{I}$ ($i=1,\ldots,a$), we define a binary vector  $z_{I,i}:P^m\rightarrow \{0,1\}$ such that:
		\begin{itemize} 
			\item $z_{I,i}(p)=1$, if  (i)  $p$ is the indicating place corresponding to $\Pi_l$, i.e., $p=p_l^{ind}$; (ii) $\Pi_l$ appears in $\varphi_{I,i}$;
   \item $z_{I,i}(p)=0$, otherwise.
		\end{itemize}

  For each proposition $\varphi_{D,j}$ of  $\varphi_{D}$ ($j=1,\ldots,b$), we define a  binary vector  $d_{D,j}:P^m\rightarrow \{0,1\}$ such that:
		\begin{itemize} 
			\item $d_{D,j}(p)=1$, if {(i) $p$ is labelled with $\pi_l$, i.e., $h^m(p)=\pi_l$; (ii) $\pi_l$ appears in $\varphi_{D,j}$;} 
   \item $d_{D,j}(p)=0$, otherwise.
		\end{itemize}


{ For subspecification  $\varphi_{F}$, we define a  binary vector  $g_{F}:P^m\rightarrow \{0,1\}$ such that:
		\begin{itemize} 
			\item $g_{F}(p)=1$, if  (i)  $p$ is the indicating place corresponding to $\Pi_l$ (resp., $h^m(p)=\Pi_l$); (ii) $\neg\Pi_l$ or $\neg\pi_l$ appears in $\varphi_{F}$;
   \item $g_{F}(p)=0$, otherwise.
		\end{itemize}
}

\begin{proposition}\label{Mtg}
Let $M_{tg}\in\mathcal{M}_{basis}$ be the solution  of the following ILP:
 \begin{equation}\label{ILP}
		\begin{aligned}
			\text{min} \quad & q(M_{tg}) \\
			\text{s.t.} \quad & z_{I,i}\cdot M_{tg}\geq 1,~i=1, \ldots, a,\\
            & d_{D,j}\cdot M_{tg}\geq 1, j=1, \ldots, b,\\
            & g_{F}\cdot M_{tg}\leq 0.\\
		\end{aligned}
	\end{equation}
 Then, $M_{tg}$ is a target marking starting from  $M_{0}$ that fulfills specification $\varphi$ and have the minimal cost.
 \end{proposition} 	

  {\textit{Proof:} 
  The first constraint of Eq.~\eqref{ILP} imposes that the indicating places that corresponding to $\varphi_{I,i}$ ($i=1,\ldots,a$) are non-empty at marking $M_{tg}$, which consequently imposes that specification  $\varphi_{I,i}$ is fulfilled at $M_{tg}$. The second constraint of Eq.~\eqref{ILP}
  enforces that places that
  corresponding to $\varphi_{D,j}$ ($j=1,\ldots,b$) are non-empty at marking $M_{tg}$, i.e., specification  $\varphi_{I,i}$ is fulfilled at $M_{tg}$. Finally, the third constraint of Eq.~\eqref{ILP} enforces that the indicating places that corresponding  to $\varphi_{F}$ is empty at marking $M_{tg}$, which  indicates that specification  $\varphi_{F}$ is fulfilled at $M_{tg}$, i.e., no agent has visited the regions corresponding to $\varphi_{F}$. Therefore, $M_{tg}$ is a target marking that meets the specification $\varphi$ and has minimal cost.} \hfill $\blacksquare$

 {Note that since only a unique path which has the minimal cumulative 
 cost is} preserved  between every basis marking pair during the construction of the EBRG (i.e., each basis marking in $\mathcal{B}$  has only one upstream basis marking, except for the initial marking), the path  from $M_0$ to  $M_{tg}$ is unique and has the minimal cost, which  can be easily found by the backtracking method. The entire planning procedure for Problem ~\ref{problem1} is shown in Algorithm~2.

 \begin{algorithm}[!htpb]\scriptsize
 		\SetAlgoLined 
 		\caption{Planning algorithm for Problem \ref{problem1}}
 		\KwIn{TAMP-PN $Q$, a Boolean specification $\varphi$ in form (1)}
 		\KwOut{Optimal sequence $\sigma^\star$ and the for Problem 1 and cost $q^\star$}
		{Construct the simplified TAMP-PN $Q^s$ of $Q$ according to Definition 1\\
  Construct the monitored TAMP-PN $Q^m$ of $Q^s$ according to Definition 2\\
   Construct the EBRG $\mathcal{B}$ of $Q^m$ according to Algorithm 1}\\
   {Compute the target marking $M_{tg}$ by Eq.~\eqref{ILP}\\}

   \uIf{ $\nexists M\in\mathcal{M}_{basis}$ such that $M=M_{tg}$} 
   {\textbf{return} There is no solution to Problem \ref{problem1} \\
   }
      \Else{
      Construct the transition sequence $\sigma_m^\star$  from  $M_0$ to  $M_{tg}$ in $Q^m$ by the backtracking method\\
      $\sigma^\star\leftarrow\mathcal{F'}(\sigma_m^\star)$\\
         \textbf{return} $\sigma^\star$\ with cost $q^\star\leftarrow w^T\cdot y_{\sigma^\star}$}  
 \end{algorithm}	

{ \begin{theorem}
 For a TAMP-PN $Q$, a Boolean specification $\varphi$ in form (1), Algorithm~2 provides an optimal solution for the task and motion planning of autonomous systems formulated in  Problem \ref{problem1}.      
\end{theorem}\label{Optimal}

\textit{Proof:} According to Proposition \ref{Mtg}, if there exists a target marking $M_{tg}\in\mathcal{M}_{basis}$ that satisfies ILP (4), then there must exist a unique path from $M_0$ to $M_{tg}$ that fulfills the specification $\varphi$. Then there exists an optimal sequence $\sigma_m^\star$ in $Q^m$ to Problem \ref{problem1}, which consequently implies that   $\sigma^\star=\mathcal{F}'(\sigma_m^\star)$ is an optimal solution to Problem \ref{problem1}. 
\hfill $\blacksquare$
  
}

\section{Experiment Results}\label{5}
{This section presents a series of experiments to  illustrate the effectiveness and scalability of the developed method. 
In subsection \ref{realcase}, we apply our proposed method within the context of a real manufacturing facility. Then,  the scalability and computational efficiency of the proposed method are illustrated and compared with some existing methods in subsection\ref{scalability}. All simulations are implemented in MATLAB on a computer running Windows 10 operating system, equipped with an Intel Core i5 2.3GHz CPU and 32 GB of RAM.} 


\subsection{Case study}\label{realcase}
{
Consider the floor diagram of a seafood processing plant shown in Fig. 2. Due to the prolonged operation under extreme conditions (high and low temperatures) in the drying and freezing rooms, the machines in theses rooms are highly susceptible to damage.
{Two UGVs are deployed}, capable of performing specific tasks such as fault detection and maintenance of equipment in the drying and freezer rooms, and conducting safety inspections of various storage areas. Given the damp environment of seafood processing plants, to enhance the agents' lifetime and prevent potential cross-contamination from food residues, each agent must proceed to a disinfection area for cleaning and sterilization  after completing tasks.}

{The planar map can be discretized into an $8\times11$ grid map using a raster method and  modeled by a PN model. The working areas are distinguished by different colors: raw material warehouse (green), freezer room (blue), drying room (red), finished product warehouse (pink), materials packaging room (purple), and disinfection room (yellow). The  set of regions of interest is $\mathbf{\Lambda}=\{\Lambda_1,\ldots,\Lambda_{10}\}$, where $\Lambda_1$ denotes disinfection room 1, $\Lambda_2$ denotes  materials packaging room, $\Lambda_3$ denotes drying room 1, $\Lambda_4$ denotes drying room 2, $\Lambda_5$ denotes raw material warehouse 1, $\Lambda_6$ denotes finished product warehouse, $\Lambda_7$ denotes disinfection room 2, $\Lambda_8$ denotes freezer room 1, $\Lambda_9$ denotes raw material warehouse 2, $\Lambda_{10}$ denotes freezer room 2. The set of atomic propositions 
is $\mathbf{\Pi}=\{\Pi_1,\Pi_2,\cdots,\Pi_{10},\pi_1,\pi_2,\cdots,\pi_{10}\}$.



\begin{table*}[htbp]\tiny
	\centering
	\caption{ Simulation results for various numbers $k$ of agents on fixed size of $20\times20$ grid environment.}
        \label{table1}
\begin{tabular}{|c|c|c|c|c|c|c|c|c|c|c|}
\hline
\multirow{2}{*}{$k$} & \multicolumn{2}{c|}{ILP\cite{Mahulea2017}} & \multicolumn{3}{c|}{SAA\cite{simulated-annealing}} & \multicolumn{3}{c|}{Our approach} \\ 
\cline{2-9} & \multicolumn{1}{c|}{$cost$} & $time [s]$  & $Bcost$ & $Gap(\%)$  & $time [s]$ & $cost$ &  $| \mathcal{M}_{basis}|$    & $time[s]$ \bigstrut\\
		\hline
		1     & 50    & 54.527  & 53 & 6.000 & 4.010 & 50 & 122        & 0.006 \bigstrut\\
		\hline
		2     & 67     & 70.454 & 67 & 0 & 3.981 & 67 & 451      & 0.014 \bigstrut\\
		\hline
		3     & 65    & 61.661 & 67 & 3.076 & 3.821 & 65 & 1319     & 0.015 \bigstrut\\
		\hline
		4     & 55    & 57.652 & 55 & 0 & 3.688 
       & 55 & 3521      & 0.023 \bigstrut\\
		\hline
		5     & 48    & 51.986 & 48 & 0 & 4.379 & 48 & 9852    & 0.031  \bigstrut\\
		\hline
		6     & 39    & 44.527 & 39 & 0 & 3.736 
 & 39 & 14785   & 0.137 \bigstrut\\
		\hline
		7     & 44    & 48.083 & 44 & 0 & 3.969 & 44 & 28751  & 0.232 \bigstrut\\
		\hline
		8     & 29     & 23.563 & 29 & 0 & 3.847 & 29 & 35570  & 0.342 \bigstrut\\
		\hline
		9     & 30    & 31.986 & 30 & 0 & 3.977 & 30 & 97183   & 0.731 \bigstrut\\
		\hline
	\end{tabular}%
	\label{tab:addlabel}%
\end{table*}%

\begin{table*}[htbp]\tiny
	\centering
	\caption{ Simulation results for various sizes $W$ of environments with $k=3$.}
        \label{table2}
	\begin{tabular}{|c|c|c|c|c|c|c|c|c|c|c|}
\hline
\multirow{2}{*}{$W$} & \multicolumn{2}{c|}{ILP\cite{Mahulea2017}} & \multicolumn{3}{c|}{SAA\cite{simulated-annealing}} & \multicolumn{3}{c|}{Our approach} \\ 
\cline{2-9} & \multicolumn{1}{c|}{$cost$} & $time [s]$ & $Bcost$ & $Gap(\%)$  & $time [s]$ &  $cost$ & $|\mathcal{M}_{basis}|$     & $time [s]$ \bigstrut\\
		\hline
		$10\times10$ & 21    & 1.824 & 21 & 0 & 2.652 & 21 & 4577     & 0.023 \bigstrut\\
        \hline
		$15\times15$ & 27    & 1.434 & 27 & 0 & 3.098 & 27 & 4973    & 0.025 \bigstrut\\
		\hline
		$20\times20$ & 33    & 15.025 & 33 & 0 & 3.952 & 33 & 6658     & 0.027 \bigstrut\\
        \hline
		$25\times25$ & 43   & 424.169 & 50 & 16.279 & 12.304 & 43 & 6857      & 0.027 \bigstrut\\
		\hline
		$30\times30$ & 57    & 1508.387 & 57 & 0 & 33.187 & 57 & 3006    & 0.012 \bigstrut\\
        \hline
		$35\times35$ & 77    & 2835.534 & 84 & 9.091 & 83.887 & 77 & 4357      & 0.021 \bigstrut\\
		\hline
		$40\times40$ & -    & - & 143 & 0 & 197.318 & 143 & 6028   & 0.035 \bigstrut\\
        \hline
		$45\times45$ & -   & - & 128 & 0 & 405.771 & 128 & 8044      & 0.034 \bigstrut\\ 
		\hline
		$50\times50$ & -    & - & 156 & 0 & 698.959 & 156 & 6021  & 0.030 \bigstrut\\
		\hline
	\end{tabular}%

\end{table*}%


\begin{figure}[!htbp]
		\centering
		\includegraphics[scale=0.2]{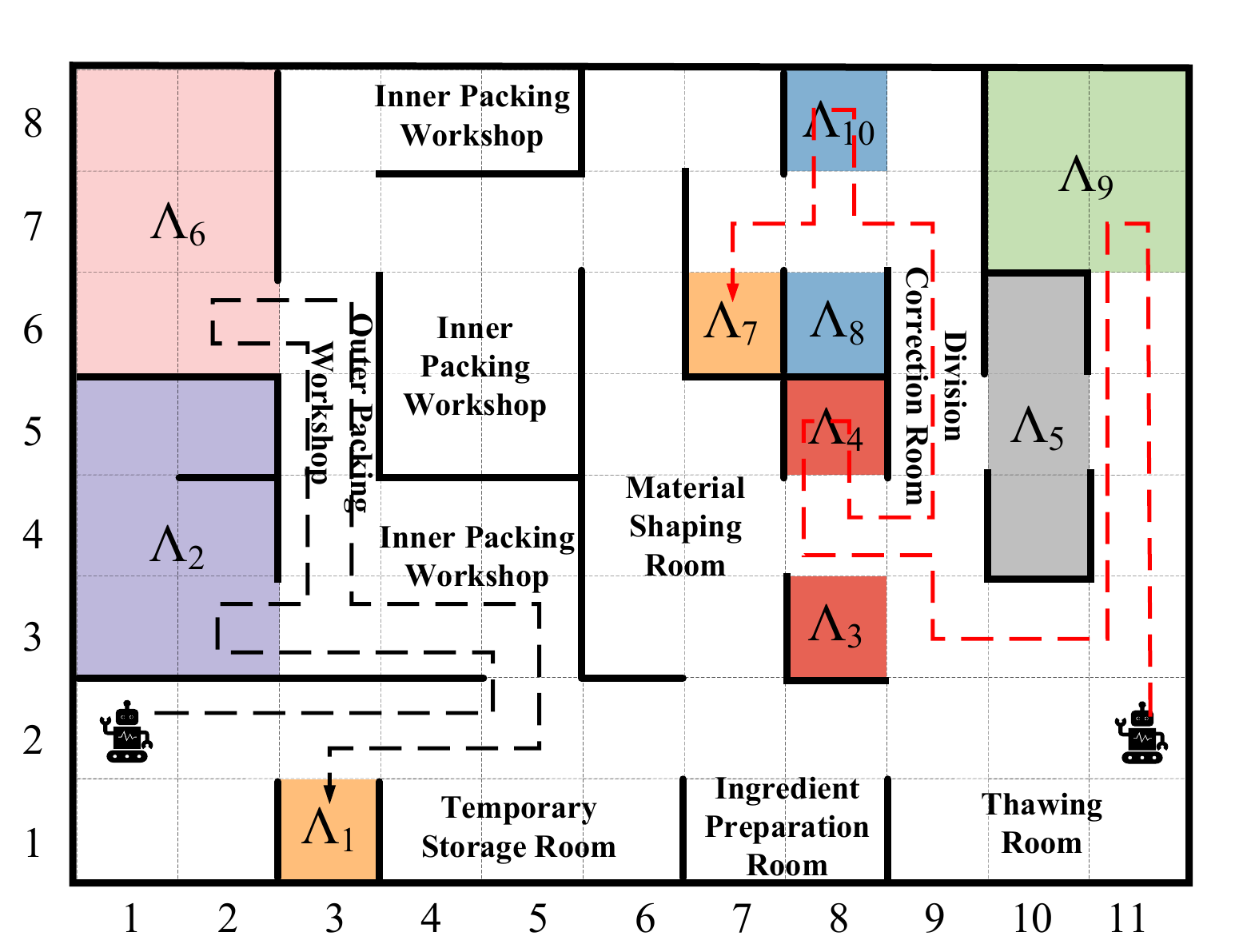}
		\caption{Planning results for task $\varphi_1$.}		
	\end{figure}

Due to ongoing maintenance in raw material warehouse 1, access to this area is prohibited and this area is marked in color gray. { This requirement can be represented by $\neg\Pi_5$.} In this situation, UGVs are required to conduct daily safety inspections of drying room 2, raw material warehouse 2, freezer room 2, finished product warehouse, and the materials packaging room, { which can be represented by $\Pi_4\wedge\Pi_9\wedge\Pi_{10}\wedge\Pi_2\wedge\Pi_6$.} Finally, all UGVs need to undergo cleaning and sanitation, which can be represented by $\pi_1\wedge\pi_7$. The task requirements can be represented by the following Boolean specification:
{\begin{equation*}	\varphi_1=\Pi_2\wedge\Pi_4\wedge(\neg\Pi_5)\wedge\Pi_6\wedge\Pi_9\wedge\Pi_{10}\wedge(\pi_1\wedge\pi_7).
	\end{equation*}

{The planning results for $\varphi_1$ are shown in Fig. 2.
The UGV following the red trajectory first conducts safety inspections in raw material warehouse 2, drying room 2, and freezer room 2. Due to the inaccessible of raw material warehouse 1, the UGV detours around it when moving to drying room 2. Finally, it proceeds to disinfection room 2  for sanitation. The UGV following the black trajectory chooses to conduct safety inspections in the materials packaging room and then the finished product warehouse. Finally, it undergoes sanitation in disinfection room 1. The obtained sequences (i.e., trajectories) are optimal whose total cost of the movements of the team is 47. The computation time for the online planning is 0.031 seconds.

\subsection{Scalability results}\label{scalability}

{ In this subsection we increase the number of agents, the size of the environment {and the number of atomic propositions} to test the scalability of our approach compared with the ILP method in \cite{Mahulea2017} and the simulated annealing algorithm (SAA) in \cite{simulated-annealing}. The parameters for the SAA are set as the same with \cite{simulated-annealing}, including number of neighborhood searches, initial temperature, final temperature, and temperature attenuation coefficient. For every experiment, The number of grids with atomic proposition is uniformly chosen between 2 and 10 and the distribution of grids with atomic propositions
also follows uniform distribution. The Boolean specification is generated randomly by using Boolean operators (conjunction, disjunction, and negation). In all environments considered in this subsection, it is assumed that the cost of traversing different grids is the same for all agents. The data presented in the tables represents the mean computation time following 20 runs.  




\begin{table*}[htbp]\scriptsize
	\centering
	\caption{ Simulation results for various numbers $A$ of atomic propositions on size of $20\times20$ grid environment with $k=3$.}
        \label{table3}
\begin{tabular}{|c|c|c|c|c|c|c|c|c|c|}
\hline
\multirow{2}{*}{$A$} & \multicolumn{2}{c|}{ILP\cite{Mahulea2017}} & \multicolumn{3}{c|}{SAA\cite{simulated-annealing}} & \multicolumn{3}{c|}{Our approach} \\ 
\cline{2-9} & \multicolumn{1}{c|}{$cost$} & $time [s]$  & $Bcost$ & $Gap(\%)$  & $time [s]$ & $cost$ &  $| \mathcal{M}_{basis}|$     & $time[s]$ \bigstrut\\
		\hline
		4     & 24    & 15.067  & 24 & 0 & 1.325 & 24 & 195        & 0.004 \bigstrut\\
		\hline
		5     & 29     & 24.528 & 29 & 0 & 2.476 & 29 & 738      & 0.016 \bigstrut\\
		\hline
		6     & 35    & 41.423 & 35 & 0 & 3.848 & 35 & 1980      & 0.020 \bigstrut\\
		\hline
		7     & 46    & 49.916 & 49 & 6.522 &  5.968
       & 46 & 4456      & 0.063 \bigstrut\\
		\hline
		8     & 51    & 58.295 & 57 & 11.764 & 7.027 & 51 & 15736    & 0.145  \bigstrut\\
		\hline
		9     & 65    & 63.241 & 71 & 9.231 & 7.526 
 & 65 & 31472   & 0.256 \bigstrut\\
		\hline
		10     & 73    & 73.587 & 80 & 9.589 & 8.933 & 73 & 82905  & 0.686 \bigstrut\\
		\hline
		11     & 79     & 80.017 & 86 & 8.860 & 9.225 &79 & 248832  & 2.430 \bigstrut\\
		\hline
		12     & 80    & 86.244 & 92 & 15.000 & 10.893 & 80 & 695328   & 7.323 \bigstrut\\
		\hline
	\end{tabular}%
	\label{tab:addlabel}%
\end{table*}%

For ease of statistics and presentation, the scalability tests of the number of agents, the size of the environments {and the number of the atomic propositions are shown in Tables \ref{table1} , \ref{table2} and \ref{table3}, respectively}. Here, $k$ is the number of agents, $W$ is the map size, { $A$ is the number of atomic propositions}, {$cost$ is the total cost of all agent movements},  $Bcost$ is the best cost founded by the SAA,
$Gap(\%)$ represents the relative tolerance to the optimal cost, which is defined as {$|Bcost-cost|/cost$}, { and \textit{time} denotes the computation time of ILP, SAA, and the online planning, respectively. Additionally, we use “-”  to indicate that ILP fails to obtain a solution due to high computation times ($>$10,000 seconds).} 
{We use $|\mathcal{M}_{basis}|$ to represent the number of basis markings of the BRG $\mathcal{B}$.}
{It should be noted that due to its main principles, our approach always leads to an optimal solution. Furthermore
concerning the complexity aspects, the numerical effort of our approach depends linearly on the size of the BRG.}

{\textit{Comparison results for different numbers of agents:} 
In this experiment, the size of the environment is fixed at $20\times20$ and the number of agents is gradually increased from 1 to 9. 
The results in Table \ref{table1} indicate that both ILP and our method can effectively provide optimal solutions. The solution time of the ILP method is mainly affected by the maximal number of moving steps, which decrease gradually  when  the number of agents increases. Furthermore, as a meta-heuristic method, SAA is able to provide planning strategies faster than the ILP method and exhibits good solution quality in the arithmetic examples. The experimental results show that the solution time of the SAA algorithm is independent of the number of agents $k$.
For our approach, the number of agents also has a slightly effect on the computation time. 
This is due to the fact that our approach removes the most burdensome part offline, while the online planning procedure is performed based on a compact representation of the state space which further reduces the computational complexity. As a conclusion,  the experiment results indicate that the growth in the number of agents has a small impact on the computation time for all methods, while our approach is the most efficient one. 

}

\textit{Comparison results for different size of environments:} 
In this test, the size of the environment is increased from $10\times10$ to $50\times50$, while the number of agents is fixed at 3. The results in Table \ref{table2} indicate that the effects of the size
of the environment on the computation time of the ILP method and SAA method are much   larger than our method. Particularly, the ILP method cannot provide a solution within 10000 seconds when the size of the environment is larger than or equal to $40\times40$.  This is due to the fact that the size of the ILP model increase significantly when the size of the environment increases. In addition, the computation time of SAA method also increases when the  size of the environment increases. Our approach can provide an optimal solution within 0.04s for all the tested cases.
The main reason is that our approach abstracts the irrelevant (empty proposition) regions during the offline phase. 
 As a conclusion,  the experiment results indicate that our method
 has significantly better scalability w.r.t the size of the environment than the existing methods.

 \textit{Comparison results for different number of atomic propositions:} In this experiment, the number of atomic propositions { (i.e., complexity of the tasks)}  is increased with the environment size of $20\times20$ and the number of agents is fixed at 3. The results in Table \ref{table3} indicate that the increase in the number of atomic propositions has { some effects on all methods. The solution time of the ILP method increases slightly when the  number of atomic propositions increases, while the solution quality of the SAA method decreases significantly due to the increase solution space. Although the number of basis markings $|\mathcal{M}_{basis}|$ increases when the  number of atomic propositions increases, the online planning is still very efficient for our approach.}
}



\section{Conclusion}\label{conclusion}
This paper addresses the TAMP for autonomous systems with high-level tasks based on Petri nets. First, we develop some simplification methods to reduce the model size. Then, by constructing the basis reachability graph of the simplified model, we remove the most burdensome part of the planning procedure offline. Next, an efficient optimal planning method is proposed to find the optimal task and path plans.  Numerical experiments demonstrate that our method has a better scalability than the existing approaches on the size of the environment and the quantity of agents. In future work, our aim is to extend the Boolean specifications to requirements for specified agent and study TAMP by considering special environments, such as the presence of sensor observations in the environment or uncertain environments.




\bibliographystyle{IEEEtran}

\bibliography{cas-refs}

\end{document}